\documentclass[a4paper]{aipproc}

\layoutstyle{6x9}
  
\begin{document}

\title{Meson Dynamics and the resulting ``3-Nucleon-Force'' diagrams:\\
      Results from a simplified test case}


\author{L. Canton}{
  address={INFN and University of Padova, Padova, Italy, I-35131}
}

\iftrue
\author{T. Melde}
{
  address={Institute for Theoretical Physics,
University of Graz, 8010 Graz, Austria},
  email={thomas.melde@kfunigraz.ac.at},
  homepage={http://physik.kfunigraz.ac.at/~thm/home.html}
}

\author{J.P. Svenne}{
  address={WITP and University of Manitoba, Winnipeg, MB, Canada R3T 2N2},  
}
\fi

\copyrightyear  {2001}

\begin{abstract}
A simplified $1D$ (one-dimensional) model for a generalized $3N$ 
system is considered, as a testing ground for the explicit treatment 
of the meson dynamics in this system. We focus attention on 
the irreducible diagrams generated by the pion dynamics in the $3N$ 
system, and in particular to a new type of three-nucleon force 
discussed recently in the literature, and generated by the 
one-pion-exchange mechanism in presence of a nucleon-nucleon 
correlation. It is found that these new terms in the simplified model 
have an approximately $30\% $ effect compared to the standard three-nucleon force 
terms in a `Triton' binding energy calculation. It is suggested that 
this effect should also not be ignored in realistic calculations.
\end{abstract}

\date{\today}

\maketitle

\section{Introduction}

An important aspect in modern three-nucleon dynamics is the
explicit inclusion of mesonic degrees of freedom that can not be
described by conventional two-nucleon potentials. These
additional effects can be described by three-nucleon forces (3NF),
which describe the addition of possible irreducible
mesonic contributions.
The construction and implementation of these terms is not
trivial and care has been taken to avoid double-contributions.
The standard
3NF approaches turned out to be able to account for the
underbinding of the three-nucleon bound state. However, so far they
could not fully explain the puzzle of the vector analyzing powers in
nucleon-deuteron scattering at low energies.\\
Recently, starting from the rigorous four-body theory of 
Grassberger-Sandhas \cite{GS67} and Yakubovsky \cite{Yak67}, 
a new method to describe the coupled $NNN-\pi NNN$ system
has been developed \cite{Can98}. In a subsequent paper,
an approximation scheme for this new method was derived by
the authors \cite{CMS01}. It is based on reasonable physical and 
mathematical approximations and a procedure to freeze-out
the pion channel.\\
After the approximations were performed and the pionic channel was 
projected out, it
was shown that the residual pion-dynamics produced correction terms 
to the standard Faddeev-Alt-Grassberger-Sandhas $3N$ equation. 
These correction terms can be interpreted as three-nucleon force 
diagrams (3NF). The approach developed
has the distinctive feature that these irreducible diagrams 
naturally appear in the Faddeev equation, and the corresponding 
3NF is generated consistently with the 3N dynamical equation 
used for the actual calculations. To our knowledge, this level 
of consistency has here been obtained for the first time.\\
A simplified 
one-dimensional model was developed, which describes in first approximation
the dynamics of the three-nucleon system \cite{Mel01}. It is based on 
a spinless one-dimensional scattering theory with a potential 
which is the $1D$ analogue of the standard Malfliet-Tjon potential. 
The simplicity 
of the model allowed a straightforward investigation of the effects 
due to the residual pion dynamics. 

\section{The simplified test case}

An approximation scheme for reducing the $\pi -NNN$ system to a 
tractable set of equations has been recently developed \cite{CMS01}.
At the lowest order, the residual effects of the pion dynamics result 
in irreducible corrections to the driving term of the Lovelace type 
$3N$ equation. The driving term now reads
\begin{equation}
   	Z_{ab}=Z_{ab}^{AGS}+
	Z_{ab}^{\pi}
\end{equation}
where $a,b$ are the standard Faddeev labels.
The first term has the usual structure of an AGS-type driving term, 
namely it is vanishing for $a=b$, while for $a\ne b$ it is
\begin{equation}
    Z_{12}^{AGS}=\left\langle {N_1\left( {N_2N_3} \right)} \right|
    g_0\left| {N_2\left( {N_1N_3} \right)} \right\rangle 
\end{equation}
and the other term corresponds to the correction 
that comes from the residual pion dynamics.  Another type of 
correction term results from projecting out the pion channel. It describes 
the intermediate propagation of a pion under the presence of a 
correlated three nucleon cluster. In this study such an additional 
correction is not taken into account.\\
To the lowest order, the correction term $Z_{ab}^{\pi}$ contains two 
topologically different contributions, for $a\ne b$, 
\begin{equation}
    Z_{12}^{\pi}=\left\langle {N_1\left( {N_2N_3\pi } \right);
    N_1N_2\left( {N_3\pi } \right)} \right|\tau _{\left( {N_3\pi } \right)}
    \left| {N_2\left( {N_3N_1\pi } \right);N_1N_2\left( {N_3\pi } \right)}
    \right\rangle 
\end{equation}
and, for $a=b$,
\begin{equation}
    Z_{11}^{\pi}=\left\langle {N_1\left( {N_2N_3\pi } \right);
    N_1\left( {N_2N_3} \right)\pi } \right|\tau _{\left( {N_2N_3} \right)}
    \left| {\left( {N_1\pi } \right)\left( {N_2N_3} \right);N_1N_3
    \left( {N_1\pi } \right)} \right\rangle 
\end{equation}
\begin{displaymath}
+\left\langle {\left( {N_1\pi } \right)\left( {N_2N_3} \right);
N_1\left( {N_2N_3} \right)\pi } \right|\tau _{\left( {N_2N_3} \right)}
\left| {N_1\left( {N_2N_3\pi } \right);N_1N_3\left( {N_1\pi } \right)} 
\right\rangle 
\end{displaymath}
The static approximation of the first term corresponds to a correction 
term usually attributed to a three-nucleon force diagram of the 
Fujita-Miyazawa type \cite{FM57}. The second term corresponds to 
a topologically 
different type of correction term that describes the propagation of an 
intermediate pion under the presence of a correlated two nucleon 
cluster. Furthermore, it should be noted that the first correction term
has contributions to the off-diagonal elements of the driving 
term only. On the other hand, the second type has contributions to 
the diagonal elements of the driving term only, which in the standard 
AGS driving term are always zero. \\
A simple test model was designed from a one-dimensional scattering 
theory describing symmetric spinless particles.
The strength parameters of the Malfliet-Tjon type potential were chosen 
in a way to reproduce the deuteron binding energy. 
The potential also shows a similar spatial behaviour as observed in 
more realistic nuclear potentials, which makes it a good candidate for 
our test-calculation. The t-matrices are described by the Unitary Pole 
Approximation (UPA), which is expected to be a good approximation due 
to the simplicity of the toy-model. The `deuteron' binding energies are 
found using a 
sturmian procedure. Subsequently we calculated the `triton' binding energy 
of this system without the correction terms and found a value of 
$-7.28MeV$. \\
In the calculation the first type of correction term was included in a static 
approximation which allowed the interpretation of 
the dynamical terms as $3NF$. They were described by the 
one-dimensional equivalent of the 
Tucson-Melbourne or Brasil type contact term \cite{TMF,Brasil}. The absence 
of spin in 
our model reduces all other terms in the Tucson-Melbourne/Brasil $3NF$ to 
identically zero. The contact term of the TM/B-$3NF$ depends on a 
freely variable parameter and the $\pi N$ scattering length.
The coefficient $a_{1}$ in the contact term, which depends on the $\pi 
N$ scattering length is defined in our simplified model by the expression
\begin{equation}
    a_{1}=-{{2\pi \mu}\over {{\hbar}^{2}}}t_{\pi N}{\left(p=0,p'=0,E=0\right)}
    \label{eq:€}
\end{equation}
where the threshold $\pi N$ t-matrix is approximated by the 
corresponding $NN$ t-matrix times an adjustable parameter $c_{\pi N}$. This 
parameter $c_{\pi N}$ can be choosen freely over a certain range to 
recover the triton binding energy. 
However, we also know that the corresponding threshold scattering lengths differ 
by a factor of $0.01$ and we choose the parameter $c_{\pi N}=0.01$ as a first guess. 
With this value we find a `triton' binding energy of $-8.17MeV$. We then 
varied the parameter $c_{\pi N}$ and found that the correction term depends 
strongly on this adjustable parameter. It is observed that it is in 
principle possible to 
adjust the parameter $c_{\pi N}$ in order to recover the 
experimental `triton' binding energy.\\
When calculating the effect of the second type of correction terms a 
delicate cancellation has to be observed in order to avoid overcounting. This 
cancellation is described by Canton and Schadow \cite{CS01b} and is 
of the type $t-v$.
Namely, we have to subtract the potential from the full t-matrix and 
it turns out that this subtraction does not result in a complete 
cancellation at low energies. 
For more details on the cancellation we refer to the original 
publications \cite{CS01b,CS01a}.
We then calculated the `triton' energy including both types of 
correction terms, where for the adjustable parameter $c_{\pi N}$ a value 
of $0.01$ was chosen, thereby obtaining $-8.48MeV$ for the model 
`triton' binding energy. It should be noted that 
the effect of the `new' correction terms in respect to the 
Fujita-Miyazawa type terms is approximately $30\% $. If we use 
the parameter $c_{\pi N}$ that was already adjusted to recover the `triton' 
binding energy, then the new type of correction term leads to an over binding. 
It is therefore important to fit the adjustable parameter $c_{\pi N}$ with both 
types of correction terms present.

\section{Conclusions}

We developed a model that included one pion degree of freedom 
explicitly in an AGS type three body equation. It was shown that the 
effect of the pion dynamics resulted in correction terms in the 
driving terms of the two-cluster equations.
We calculated the explicit effects of these terms, which could be 
interpreted as
3NF terms, on the `triton' binding energy. In a simplified model calculation 
we show that the correction terms can account for an underbinding
of the `three-nucleon' bound state. However, a new type of correction 
term also
has an approximately $30\% $ effect on the `triton' binding energy. A 
third topological type of correction term still needs to be 
investigated, but it is expected that the effect of this term is not 
as important.\\
The calculations we presented were done only with a rather simple 
model. However, one might expect that
the new type of correction terms could have an effect also in more
realistic calculations.
This suggestion has been recently confirmed by Canton and Schadow
in a more realistic study \cite{CS01b,CS01a} on the $nd$ vector analyzing 
powers, where it was shown that a new 3NF term of tensor structure
can be generated by these new correction terms, and could indeed hold 
the key to solve the $A_{y}$-puzzle.
In conclusion, a new way of handling the
pion dynamics in $3N$ calculations has been developed. Calculations 
suggest that a newly developed type of 3NF should also be included 
in realistic three-nucleon calculations.

\begin{theacknowledgments}
L.C. thanks for support from the Italian MURST-PRIN Project ``Fisica 
Teorica del Nucleo e dei Sistemi a Piu Corpi''. T.M acknowledges 
support from the University of Manitoba and the conference organizing committee
for waiving the conference fee. J.P.S. acknowledges support from the 
Natural Science and Engineering Research Council of Canada. The 
authors also would like to thank the Institutions of INFN (Padova), 
University of Manitoba, and Universita di Padova for hospitality 
during reciprocal visits.
\end{theacknowledgments}



\end{document}